\documentstyle[12pt]{article}

\begin{document}

%%-move to normal A4-%%
\hoffset = -0.3truecm
\voffset = -1.1truecm

\title{\bf
STATIC MONOPOLES AND THEIR ANTI-CONFIGURATIONS\footnote{USM Preprint June 2004; Hep-th/0406094; Accepted for publication in the International Journal of  Modern Physics A}}

\author{
{\bf Rosy Teh\footnote{e-mail: rosyteh@usm.my} and Khai-Ming Wong}\\
{\normalsize School of Physics, Universiti Sains Malaysia}\\
{\normalsize 11800 USM Penang, Malaysia}}

\date{June 2004}
\maketitle

\begin{abstract}
Recently, we have reported on the existence of some monopoles, multimonopole, and antimonopoles configurations. In this paper we would like to present more monopoles, multimonopole, and antimonopoles configurations of the magnetic ansatz of Ref.\cite{kn:9} when the parameters $p$ and $b$ of the solutions takes different serial values. These exact solutions are a different kind of BPS solution. They satisfy the first order Bogomol'nyi equation but possess infinite energy. They can have radial, axial, or rotational symmetry about the z-axis. We classified these serial solutions as (i) the multimonopole at the origin; (ii) the finitely separated 1-monopoles; (iii) the screening solutions of multimonopole and  (iv) the axially symmetric monopole solutions. We also give a construction of their anti-configurations with all the magnetic charges of poles in the configurations reversed. Half-integer topological magnetic charge multimonopole also exist in some of these series of solutions.
\end{abstract}

%%-main body of paper-%%
%%-self numbering sections-%%

\section{Introduction}
The SU(2) Yang-Mills-Higgs (YMH) field theory, with the Higgs field in the adjoint representation are known to possess both the magnetic monopole and multimonopole solutions. The 't Hooft-Polyakov monopole solution, \cite{kn:1} with non zero Higgs mass and Higgs self-interaction, belongs to the category of solutions which are invariant under a U(1) subgroup of the local SU(2) gauge group. This numerical monopole solution of unit magnetic charge is spherically symmetric. 
In general, configurations of the YMH field theory with a unit magnetic charge are spherically symmetric \cite{kn:1}-\cite{kn:3}. However we would like to point out that we have presented a unit magnetic charge configuration in Ref.\cite{kn:12} that do not even possess axial symmetry but only mirror symmetry. 

In the limit when the Higgs mass and the Higgs self-interaction tend to zero with the vacuum expectation value non vanishing, the Higgs field becomes massless and is non self-interacting. This model, with non-zero vacuum expectation is known as the Bogomol'nyi-Prasad-Sommerfield (BPS) limit as exact solutions can be obtained by solving the first order Bogomol'nyi equations \cite{kn:3}, \cite{kn:6}. These BPS solutions possess minimal energies.
Exact BPS multimonopole configurations with magnetic charges greater than unity and possessing axial and mirror symmetries were reported in the early 80's \cite{kn:4} and it has been shown that these solutions cannot possess spherical symmetry \cite{kn:5}.

However when the Higgs potential is finite only numerical monopole solutions \cite{kn:1}, \cite{kn:2} and numerical axially symmetric multimonopole solutions are known \cite{kn:7} . Asymmetric multimonopole solutions with no rotational symmetry also exist \cite{kn:8}, however these solutions are numerical solutions even in the BPS limit.

Axially symmetric monopoles-antimonopoles chain solutions which do not satisfy the Bogomol'nyi condition are also constructed numerically. These non-Bogomol'nyi solutions exist both in the limit of a vanishing Higgs potential as well as in the presence of a finite Higgs potential. BPS axially symmetric vortex rings solutions have also been constructed numerically \cite{kn:7}. 

Recently we have shown the existence of a different kind of BPS static monopole solutions \cite{kn:9}. Other than the Wu-Yang type 1-monopole, these configurations possess at most axial symmetry and they represent different combinations of monopoles, multimonopole, and antimonopoles, with mirror symmetry about the z-axis. 

In this paper we would like to present more monopoles, multimonopole, and antimonopoles configurations of the magnetic ansatz of Ref.\cite{kn:9} when the parameters $p$ and $b$ of the solutions takes different serial values. Similarly,  these exact solutions are a different kind of BPS solutions. They satisfy the first order Bogomol'nyi equation but possess infinite energy. They can have radial, axial, or rotational symmetry about the z-axis. 

It is also our purpose in this paper to attempt to summarize all the possible monopole configurations that the magnetic anzatz of Ref.\cite{kn:9} are able to support.
We noticed that we can classified these serial monopole solutions as (i) the multimonopole at the origin; (ii) the finitely separated 1-monopoles; (iii) the screening solutions of multimonopole by antimonopoles and (iv) the axially symmetric (AS) configurations. The multimonopole in some of these series of solutions can have half-integer topological magnetic charge. The magnetic ansatz also admits isolated one-half topological charge monopoles which we have reported in a separate work \cite{kn:12}.

The fourth class of solutions are axially symmerical about the z-axis and they represent antimonopole-monopole-antimonopole (A-M-A) and vortex rings configurations. The net magnetic charge of these configurations is always negative one, whilst the net magnetic charge at the origin is always positive one for all positive integer values of the solution parameter $m$. However, when $m$ increases beyond one, vortex rings appear and the number of these rings increases proportionally with the value of $m$. The vortex rings are magnetically neutral and are located in space where the Higgs field vanishes. This work is reported in a separate paper \cite{kn:11}.

Finally in this paper, we give a construction of the anti-configuration of all possible monopoles solutions of the magnetic ansatz of Ref.\cite{kn:9} where all the magnetic charges of poles in the configurations are reversed.
This can be done by just modifying some signs in the magnetic ansatz and changing the $\phi$ winding number from one to negative one. Upon solving the Bogomol'nyi equations with the negative sign, we are able to generate exactly all the anti-configuration of all the possible monopoles solutions within the magnetic ansatz of Ref.\cite{kn:9}.  

We briefly review the SU(2) Yang-Mills-Higgs field theory in the next section. We present the magnetic ansatz and some of its basic properties in section 3. The screening of multimonopole by antimonopoles is discussed in section 4. In section 5, we discussed the configurations with only monopoles and multimonopole and no antimonopoles. In section 6, we give the construction of the anti-configuration of all possible monopoles solutions obtainable from the ansatz of Ref.\cite{kn:9}. We end with some comments in section 7.

\section{The SU(2) Yang-Mills-Higgs Theory}
The SU(2) YMH theory admits the triplet gauge fields $A^a_\mu$ which are the Yang-Mills vector fields coupled to a scalar Higgs triplets field $\Phi^a$ in 3+1 dimensions.
The index $a$ is the SU(2) internal space index and for a given $a$, $\Phi^a$ is a scalar whereas $A^a_\mu$ is a vector under Lorentz transformation. The Lagrangian in 3+1 dimensions is given by 

\begin{equation}
{\cal L} = -\frac{1}{4}F^a_{\mu\nu} F^{a\mu\nu} + \frac{1}{2}D^\mu \Phi^a D_\mu \Phi^a - \frac{1}{4}\lambda(\Phi^a\Phi^a - \frac{\mu^2}{\lambda})^2, 
\label{eq.1}
\end{equation}

\noindent where the Higgs field mass, $\mu$, and the strength of the Higgs potential, $\lambda$, are constants. The vacuum expectation value of the Higgs field is then given by $\mu/\sqrt{\lambda}$. The Lagrangian (\ref{eq.1}) is gauge invariant under the set of independent local SU(2) transformations at each space-time point.
The covariant derivative of the Higgs field and the gauge field strength tensor are given respectively by 
\begin{eqnarray}
D_{\mu}\Phi^{a} &=& \partial_{\mu} \Phi^{a} + \epsilon^{abc} A^{b}_{\mu}\Phi^{c}, ~~\mbox{and}\nonumber\\
F^a_{\mu\nu} &=& \partial_{\mu}A^a_\nu - \partial_{\nu}A^a_\mu + \epsilon^{abc}A^b_{\mu}A^c_\nu.
\label{eq.2}
\end{eqnarray}
Since the gauge field coupling constant $g$ can be scaled away, we can set $g$ to one without any loss of generality. The metric used is $g_{\mu\nu} = (-+++)$. The SU(2) internal group indices $a, b, c$ run from 1 to 3 and the spatial indices are $\mu, \nu, \alpha = 0, 1, 2$, and $3$ in Minkowski space.

The equations of motion that follow from the Lagrangian (\ref{eq.1}) are
\begin{eqnarray}
D^{\mu}F^a_{\mu\nu} &=& \partial^{\mu}F^a_{\mu\nu} + \epsilon^{abc}A^{b\mu}F^c_{\mu\nu} = \epsilon^{abc}\Phi^{b}D_{\nu}\Phi^c,\nonumber\\
D^{\mu}D_{\mu}\Phi^a &=& -\lambda\Phi^a(\Phi^{b}\Phi^{b} - \frac{\mu^2}{\lambda}).
\label{eq.3}
\end{eqnarray}
The Abelian electromagnetic field tensor as proposed by 't Hooft \cite{kn:1}, is given by 
\begin{eqnarray}
F_{\mu\nu} &=& \hat{\Phi}^a F^a_{\mu\nu} - \epsilon^{abc}\hat{\Phi}^{a}D_{\mu}\hat{\Phi}^{b}D_{\nu}\hat{\Phi}^c\nonumber\\
&=& \partial_{\mu}A_\nu - \partial_{\nu}A_\mu - \epsilon^{abc}\hat{\Phi}^{a}\partial_{\mu}\hat{\Phi}^{b}\partial_{\nu}\hat{\Phi}^c,
\label{eq.4}
\end{eqnarray}

\noindent where $A_\mu = \hat{\Phi}^{a}A^a_\mu$, the Higgs field unit vector $\hat{\Phi}^a = \Phi^a/|\Phi|$ and the Higgs field magnitude $|\Phi| = \sqrt{\Phi^{a}\Phi^{a}}$. 
The Abelian electric field is $E_i = F_{0i}$, and the Abelian magnetic field is $B_i = -\frac{1}{2}\epsilon_{ijk}F_{jk}$. 
The topological magnetic current \cite{kn:10} which is also the topological current density of the system is defined to be 
\begin{eqnarray}
k_\mu = \frac{1}{8\pi}~\epsilon_{\mu\nu\rho\sigma}~\epsilon_{abc}~\partial^{\nu}\hat{\Phi}^{a}~\partial^{\rho}\hat{\Phi}^{b}~\partial^{\sigma}\hat{\Phi}^c,
\label{eq.5}
\end{eqnarray}
and the corresponding conserved topological magnetic charge is
\begin{eqnarray}
M & = & \int d^{3}x~k_0 = \frac{1}{8\pi}\int \epsilon_{ijk}\epsilon^{abc}\partial_{i}\left(\hat{\Phi}^{a}\partial_{j}\hat{\Phi}^{b}\partial_{k}\hat{\Phi}^{c}\right)d^{3}x\nonumber\\
& = & \frac{1}{8\pi}\oint d^{2}\sigma_{i}\left(\epsilon_{ijk}\epsilon^{abc}\hat{\Phi}^{a}\partial_{j}\hat{\Phi}^{b}\partial_{k}\hat{\Phi}^{c}\right)\nonumber\\
& = & \frac{1}{4\pi} \oint d^{2}\sigma_{i}~B_i. 
\label{eq.6}
\end{eqnarray}

Our work is restricted to the static case where $A^a_0 = 0$. Hence the conserved energy of the system which is obtained from the Lagrangian (\ref{eq.1}) in the usual way, reduces for the static case to
\begin{eqnarray}
E = \int d^{3}x\left(\frac{1}{2}B^a_iB^a_i + \frac{1}{2}D_{i}\Phi^{a}D_{i}\Phi^a + \frac{1}{4}\lambda(\Phi^{a}\Phi^{a} - \frac{\mu^2}{\lambda})^2\right).
\label{eq.7}
\end{eqnarray}

\noindent Here $i, j, k$ which are the three space indices run from 1, 2, and 3. This energy  vanishes when the gauge potential, $A^a_i$ is zero or when $A^a_i$ is a pure gauge, and when $\Phi^a\Phi_a = \mu^2/\lambda$ and $D_i\Phi^a = 0$.

In the model we are considering, the Higgs field is massless and with vanishing self-interaction. Hence the Lagrangian (\ref{eq.1}) is just simply

\begin{equation}
{\cal L} = -\frac{1}{4}F^a_{\mu\nu} F^{a\mu\nu} + \frac{1}{2}D^\mu \Phi^a D_\mu \Phi^a. 
\label{eq.8}
\end{equation}
The magnitude of the Higgs field vanishes as $1/r$ at large $r$. It is in this limit that we are able to obtain explicit, exact magnetic monopoles solutions to the YMH equations. These solutions can be solved using both the second order Euler-Lagrange equations and the Bogomol'nyi equations,

\begin{equation}
B^a_i \pm D_i \Phi^a = 0. 
\label{eq.9}
\end{equation}

The $\pm$ sign corresponds to monopoles and antimonopoles respectively for the usual BPS solutions. In our case the magnetic monopoles of Ref.\cite{kn:9} as well as the solutions presented here in section 3 are solved with the $+$ sign and their anti-configurations in section 4 are solved with the $-$ sign. The configurations obtained correspond to different combinations of monopole, multimonopoles, and antimonopoles together with their anti-configurations. In this paper, a multimonopole of magnetic charge $M$ and multi-antimonopole with all its magnetic charges superimposed at one point in space is denoted by a $M$-monopole and a $M$-antimonopole respectively. 
When $M=1$, it is possible for the monopoles and antimonopoles respectively to have finite separations in space. However we failed to find $M$-monopoles and $M$-antimonopole with finite separations when $M>1$. 

\section{The Magnetic Ansatz}
We make use of the static magnetic ansatz \cite{kn:9} to solve for the monopoles solutions here. The gauge fields and the Higgs field are given respectively by 
\begin{eqnarray}
A_{\mu}^a &=& \frac{1}{r}\psi(r)\left(\hat{\theta}^{a}\hat{\phi}_\mu - \hat{\phi}^{a}\hat{\theta}_{\mu}\right)  + \frac{1}{r}R(\theta)\left(\hat{\phi}^{a}\hat{r}_{\mu} - \hat{r}^{a}\hat{\phi}_{\mu}\right)\nonumber\\
&+& \frac{1}{r}G(\theta,\phi)\left(\hat{r}^{a}\hat{\theta}_\mu - \hat{\theta}^{a}\hat{r}_{\mu}\right),\nonumber\\
\Phi^a &=& \Phi_{1}~\hat{r}^a + \Phi_{2}~\hat{\theta}^a + \Phi_3~\hat{\phi}^a.
\label{eq.10}
\end{eqnarray}

\noindent where $\Phi_1 = \frac{1}{r}\psi(r), ~\Phi_2 = \frac{1}{r}R(\theta), ~\Phi_3 = \frac{1}{r}G(\theta, \phi)$. The spherical coordinate orthonormal unit vectors, $\hat{r}_i$, $\hat{\theta}_i$, and $\hat{\phi}_i$ are defined by 
\begin{eqnarray}
\hat{r}_i &=& \sin\theta ~\cos \phi ~\delta_{1i} + \sin\theta ~\sin \phi ~\delta_{2i} + \cos\theta ~\delta_{3i},\nonumber\\
\hat{\theta}_i &=& \cos\theta ~\cos \phi ~\delta_{1i} + \cos\theta ~\sin \phi ~\delta_{2i} - \sin\theta ~\delta_{3i},\nonumber\\
\hat{\phi}_i &=& -\sin \phi ~\delta_{1i} + \cos \phi ~\delta_{2i},
\label{eq.11}
\end{eqnarray}

\noindent where $r=\sqrt{x^i x_i}$, ~$\theta=\cos^{-1}(x_3/r)$, and ~$\phi=\tan^{-1}(x_2/x_1)$. The gauge field strength tensor and the covariant derivative of the Higgs field are given respectively by
\begin{eqnarray}
F^a_{\mu\nu} &=& \frac{1}{r^2}\hat{r}^{a}\left\{\dot{R} + R \cot\theta + 2\psi - \psi^2 + G^{\phi} \csc\theta\right\}(\hat{\phi}_{\mu}\hat{\theta}_{\nu} - \hat{\phi}_{\nu}\hat{\theta}_{\mu})\nonumber\\
&+& \frac{1}{r^2}\left\{\hat{\theta}^{a} R(1-\psi) + \hat{\phi}^a G(1-\psi)\right\}(\hat{\phi}_{\mu}\hat{\theta}_{\nu} - \hat{\phi}_{\nu}\hat{\theta}_{\mu})\nonumber\\ 
&+& \frac{1}{r^2}\left\{\hat{r}^{a}R(1-\psi) + \hat{\phi}^a G(\cot\theta-R)\right\}(\hat{r}_{\mu}\hat{\phi}_{\nu} - \hat{r}_{\nu}\hat{\phi}_{\mu})\nonumber\\
&+& \frac{1}{r^2}\hat{\theta}^{a}\left\{r\psi^{\prime} + R \cot\theta - R^{2} + G^\phi \csc\theta\right\}(\hat{r}_{\mu}\hat{\phi}_{\nu} - \hat{r}_{\nu}\hat{\phi}_{\mu})\nonumber\\ 
&+& \frac{1}{r^2}\left\{-\hat{r}^aG(1-\psi) + \hat{\theta}^a(\dot{G} + RG)\right\}(\hat{r}_{\mu}\hat{\theta}_{\nu} - \hat{r}_{\nu}\hat{\theta}_{\mu})\nonumber\\
&+& \frac{1}{r^2}\left\{- \hat{\phi}^{a}(r\psi^{\prime} + \dot{R} - G^2)\right\}(\hat{r}_{\mu}\hat{\theta}_{\nu} - \hat{r}_{\nu}\hat{\theta}_{\mu}),\nonumber\\
D_{\mu}\Phi^{a} &=& \frac{1}{r^2}\left\{\hat{r}^{a}(r\psi^{\prime}-\psi-R^2-G^2) - \hat{\theta}^{a}R(1-\psi) - \hat{\phi}^{a}G(1-\psi)\right\}\hat{r}_{\mu}\nonumber\\
&+& \frac{1}{r^2}\left\{-\hat{r}^{a}R(1-\psi) + \hat{\theta}^{a}(\dot{R}+\psi-\psi^2-G^2) + \hat{\phi}^{a}(\dot{G}+RG)\right\}\hat{\theta}_{\mu}\nonumber\\
&+& \frac{1}{r^2}\left\{-\hat{r}^{a}G(1-\psi) - \hat{\theta}^{a}G(\cot\theta-R)\right\}\hat{\phi}_{\mu}\nonumber\\
&+& \frac{1}{r^2}\left\{\hat{\phi}^{a}(\psi-\psi^2+R\cot\theta-R^2+G^\phi\csc\theta)\right\}\hat{\phi}_{\mu},
\label{eq.12}
\end{eqnarray}
where prime means $\partial/\partial r$, dot means $\partial/\partial \theta$ and superscript $\phi$ means $\partial/\partial\phi$. 
The gauge fixing condition that we used here is the radiation or Coulomb gauge, $\partial^i A^a_i = 0$, $A^a_0 = 0$. 

By substituting the ansatz (\ref{eq.10}) into the equations of motion (\ref{eq.3}) as well as the Bogomol'nyi equations (\ref{eq.9}) with the positive sign, these equations can be simplified to just four first order differential equations,
\begin{equation}
r\psi^{\prime} + \psi - \psi^2 = -p,
\label{eq.13}
\end{equation}
\begin{equation}
\dot{R} + R\cot\theta - R^2 = p - b^2\csc^2\theta,
\label{eq.14}
\end{equation}
\begin{equation}
\dot{G} + G\cot\theta = 0,~~~G^\phi\csc\theta - G^2 = b^2\csc^2\theta,
\label{eq.15}
\end{equation}
where $p$ and $b^2$ are arbitary constants. Eq.(\ref{eq.13}) is exactly solvable for all real values of $p$ and the integration constant can be scaled away by letting $r\rightarrow r/c$, where $c$ is the arbitrary integration constant. In order to have solutions of $\psi$ with $(2m+1)$ powers of $r$, we write $p=m(m+1)$. Eqs.(\ref{eq.15}) are also exactly solvable and a general physical solution is
$G(\theta, \phi) = b\csc\theta \tan(b\phi)$, where $b$ is restricted to take half-integer values for $G$ to be a single value function. Therefore we can write $b=m \pm s$ where $s=0, 1, 2, 3, \dots$ and $m$ can take half-integer values. For all the solutions presented here $b$ is non zero. When $b=0$, we have the axially symmetric monopole solutions which is reported in a separate work \cite{kn:11}.

Eq.(\ref{eq.14}) is a Riccati equation and $R(\theta)$ can be exactly solved for different positive integer values of $s$, when $p=m(m+1)$ and $b=m \pm s$. For each integer value of $s$, we have a series of monopoles configurations. These series of solutions can be classified into four different classes of solutions. They are (i) the multimonopole at the origin, $r=0$; (ii) the finitely separated monopoles configurations; (iii) the screening solutions of multimonopole by antimonopoles; and finally (iv) the axially symmetric monopole-antimonopole and vortex rings solutions.

The solutions for the profile functions $\psi(r)$ and $G(\theta,\phi)$ are standard,
\begin{equation}
\psi(r) = \frac{(m+1)-m r^{2m+1}}{1+r^{2m+1}}, ~~~
G(\theta, \phi) = (m \pm s)\csc\theta \tan(m \pm s)\phi,
\label{eq.16}
\end{equation}
where we fixed $s$ to be a positive integer. For $\psi(r)$ to have integer powers of $r$, and $G(\theta,\phi)$ to be a single value function, the value of $m$ is restricted to a half-integer, where $m\geq -\frac{1}{2}$.

It is the Riccati equation (\ref{eq.14}) that give rise to the different monopoles configurations as the solution for $R(\theta)$ is non-unique. The profile function $R(\theta)$ when $b=m+s$, is given by
\begin{eqnarray}
R(\theta) &=& (m+1)\cot\theta +(s-1)\csc\theta \frac{Q^{m+s}_{m+1}(\cos\theta)}{Q^{m+s}_m(\cos\theta)},\nonumber\\
s &=& 1,~2,~3,~ \dots,~~~m = -\frac{1}{2}, 0, \frac{1}{2}, 1, \frac{3}{2}, 2, \dots, 
\label{eq.17}
\end{eqnarray}
where $Q^{m+s}_m(\cos\theta)$ is the associated Legendre function of the second kind of degree $m$ and order $m+s$.
We label the monopoles configurations of Eq.(\ref{eq.17}) as the 2s series of monopole solutions with no antimonopoles in the systems. We further subdivide the 2s solutions into the A2s series when $s$ is even and the B2s series when $s$ is odd.

The A2s configuration is that of a multimonopole at the origin, $r=0$. The B2s series is the configuration where the monopoles are finitely separated and arranged in a circle about the z-axis.

The profile function $R(\theta)$when $b=m-s$, is given by
\begin{eqnarray}
R(\theta) &=& (m+1)\cot\theta -(s+1)\csc\theta \frac{P^{m-s}_{m+1}(\cos\theta)}{P^{m-s}_m(\cos\theta)},\nonumber\\
s &=& 0, 1, 2, 3, \dots,~~~m = s+\frac{1}{2}, s+1\frac{1}{2}, s+2, \dots,
\label{eq.18}
\end{eqnarray}
where $P^{m-s}_m(\cos\theta)$ is the associated Legendre function of the first kind of degree $m$ and order $m-s$. The solution (\ref{eq.18}) is the screening solutions of a multimonopole by antimonopoles which we label as the 1s series of solutions.
Similarly, we further subdivide the 1s solutions into the B1s configurations when $s$ is zero or even and the A1s configurations when $s$ is odd. 
The axially symmetric monopoles solutions which is not discussed here is the series when $b=0$, hence the function $G(\theta, \phi)$ vanishes, and $p=m(m+1)$ \cite{kn:11}.

The energy of the system of solutions here is not finite due to the singuarity of the solutions at the origin, $r=0$. Also the vacuum expectation values of our solutions tend to zero at large $r$. Hence unlike the normal BPS solutions, the energy of our solutions is not bounded from below.

The net topological charge of the system is given by the integration of the radial component of the Abelian magnetic field over the sphere at infinity,  
\begin{equation}
M_\infty = \frac{1}{8\pi}\oint d^{2}\sigma_{i}\left.\left(\epsilon_{ijk}\epsilon^{abc}\hat{\Phi}^{a}\partial_{j}\hat{\Phi}^{b}\partial_{k}\hat{\Phi}^{c}\right)\right|_{r\rightarrow\infty}.
\label{eq.19}
\end{equation}
Hence the monopoles and antimonopoles of our solutions here can then be associated with the number of zeros of $\Phi^a$ enclosed by the sphere at infinity. 
The positions of the monopoles and antimonopoles of our solutions correspond to the zeros of the Higgs field but the multimonopole is always located at the origin of the coordinate axes where the Higgs field is singular. 
The definition for the magnetic charges as given by Eq.(\ref{eq.6}) and Eq.(\ref{eq.19}) is not affected by the fact that the magnitude of the Higgs field, $|\Phi|$, vanishes at large $r$. It only depends on the direction of the unit vector of the Higgs field, $\hat{\Phi}^a$, in internal space.

From the ansatz (\ref{eq.10}), $A_{\mu} = \hat{\Phi}^{a}A^{a}_{\mu} = 0$. Hence the Abelian electric field is always zero and the Abelian magnetic field is independent of the gauge fields $A^a_\mu$. To calculate for the Abelian magnetic field $B_i$, we rewrite the Higgs field of Eq.(\ref{eq.10}) from the spherical to the Cartesian coordinate system, \cite{kn:7}, \cite{kn:9} 
\begin{eqnarray}
\Phi^a &=& \Phi_{1}~\hat{r}^{a} + \Phi_{2}~\hat{\theta}^{a} + \Phi_3~\hat{\phi}^{a}\nonumber\\
&=& \tilde{\Phi}_1 ~\delta^{a1} + \tilde{\Phi}_2 ~\delta^{a2} + \tilde{\Phi}_3 ~\delta^{a3}
\label{eq.20}
\end{eqnarray}
\begin{eqnarray}
\mbox{where}~~~\tilde{\Phi}_1 &=& \sin\theta \cos \phi ~\Phi_1 + \cos\theta \cos \phi ~\Phi_2 - \sin \phi ~\Phi_3
= |\Phi|\cos\alpha \sin\beta\nonumber\\
\tilde{\Phi}_2 &=& \sin\theta \sin \phi ~\Phi_1 + \cos\theta \sin \phi ~\Phi_2 + \cos \phi ~\Phi_3
= |\Phi|\cos\alpha \cos\beta\nonumber\\
\tilde{\Phi}_3 &=& \cos\theta ~\Phi_1 - \sin\theta ~\Phi_2 = |\Phi|\sin\alpha,
\label{eq.21}
\end{eqnarray}
The Higgs unit vector is then simplified to 
\begin{eqnarray}
\hat{\Phi}^a &=& \cos\alpha \sin\beta ~\delta^{a1} + \cos\alpha \cos\beta ~\delta^{a2} + \sin\alpha ~\delta^{a3},\\
\label{eq.22}
\mbox{where},~~~\sin\alpha &=& \frac{\psi\cos\theta - R \sin\theta}{\sqrt{\psi^2+R^2+G^2}},\nonumber\\
\beta = \gamma - \phi,~~~\gamma &=& \tan^{-1}\left(\frac{\psi\sin\theta + R \cos\theta}{G}\right),
\label{eq.23}
\end{eqnarray}
and the Abelian magnetic field is found to be
\begin{eqnarray}
B_i &=& \frac{1}{r^2 \sin\theta}\left\{\frac{\partial\sin\alpha}{\partial\theta}\frac{\partial\beta}{\partial\phi} - \frac{\partial\sin\alpha}{\partial\phi}\frac{\partial\beta}{\partial\theta}\right\}\hat{r}_i\nonumber\\
&+& \frac{1}{r\sin\theta}\left\{\frac{\partial\sin\alpha}{\partial\phi}\frac{\partial\beta}{\partial r} - \frac{\partial\sin\alpha}{\partial r}\frac{\partial\beta}{\partial\phi}\right\}\hat{\theta}_i\nonumber\\
&+& \frac{1}{r}\left\{\frac{\partial\sin\alpha}{\partial r}\frac{\partial\beta}{\partial\theta} - \frac{\partial\sin\alpha}{\partial\theta}\frac{\partial\beta}{\partial r}\right\}\hat{\phi}_i.
\label{eq.24}
\end{eqnarray}
Defining the Abelian field magnetic flux as 
\begin{eqnarray}
\Omega = 4\pi M = \oint d^{2}\sigma_{i} B_i = \int B_{i}(r^{2}\sin\theta d\theta )\hat{r}_{i}~d\phi,
\label{eq.25}
\end{eqnarray}
the magnetic charge enclosed by the sphere centered at $r=0$ and of fixed radius $r_1$, is found to be 
\begin{eqnarray}
M_{r_1} &=& \frac{1}{4\pi}\int^{2\pi}_0\int^\pi_0\left.\left(\frac{\partial\sin\alpha}{\partial\theta}\frac{\partial\beta}{\partial\phi} - \frac{\partial\sin\alpha}{\partial\phi}\frac{\partial\beta}
{\partial\theta}\right) d\theta d\phi\right|_{r=r_1}.
\label{eq.26}
\end{eqnarray}
Hence the magnetic charge enclosed by the sphere of infinite radius is denoted by $M_\infty$ and the magnetic charge enclosed by the sphere of vanishing radius is denoted by $M_0$. We also denote the net magnetic charge of the surrounding antimonopoles in the 1s series of solutions by $M_{A}$.

\section{The Screening of Multimonopole}
The screening of multimonopole occurs in the 1s series of monopole solutions. This series is farther subdivided into two alternating series, the A1s series when the parameter $s$ is odd and the B1s series when $s$ is zero or even. The first two members of these two series are the A1 and the B1 solutions of Ref.\cite{kn:9} when $m$ is a natural number. The boundaries conditions of the 1s series of solutions are 
\begin{eqnarray}
\psi(r)|_{r\rightarrow \infty} &\rightarrow& -m,~~\psi(r)|_{r\rightarrow 0}\rightarrow (m+1);~~~
G(\theta, 0) = G(\theta, 2\pi)=0; \nonumber\\
R(\theta)\sin\theta|_{\theta\rightarrow 0} &\rightarrow& -(m-s),~~R(\theta)\sin\theta|_{\theta\rightarrow \pi}\rightarrow (m-s). 
\label{eq.27}
\end{eqnarray}
The difference between the A1s and B1s solutions is in the profile function, $R(\theta)$. For the A1s solutions, $R(\theta)\cos\theta|_{\theta=\pi/2} = 1$, whereas for the B1s solutions, $R(\frac{\pi}{2}) = 0$.
 
In these series, the value of $M_\infty$ can be calculated by exact integration using Maple 9. The magnetic charge $M_0$ can be obtained by the use of the approximation methods in the Maple 9 software. The parameter $m$ here can increase in steps of one-half starting from $s+\frac{1}{2}$ for each of the A1s and B1s series of solutions. 

Hence the B1 series can possess  multimonopole at $r=0$ with magnetic charge, $M_0 = 1, 2, 3, \dots$ when $m=\frac{1}{2}, 1,\frac{3}{2}, \dots$ respectively. This multimonopole is surrounded by an equal number of antimonopoles. In Ref.\cite{kn:9}, we only discussed multimonopole of even monopole charge. Here we have noticed that $m$ can take half-integer values and hence the multimonopole can also possesses odd values of monopole charge. Therefore when $m=\frac{1}{2}$, the B1 configuration is a pair of monopole and antimonopole. The monopole is at the origin, whilst the antimonopole is at the point ($\sqrt{3}, 0, 0$).

The screening antimonopoles are of monopole charge $-1$ each. The number of antimonopoles is equal to the charge of the multimonopole and they are all symmetrically arranged on a circle of radius $r = \sqrt[2m+1]{\frac{m+1}{m}}$ about the multimonopole at the origin. Hence these configurations reside in the topologically  trivial sector of the monopole solution and the net topological charge of the B1 solution is zero. 

The magnetic charge of the multimonopole at $r=0$ of the A1s solutions possess half-integer values of monopole charge when $m$ is a half-odd-integer, whereas the magnetic charge of the multimonopole of the B1s solutions are always of integer values of monopole charge. The surrounding screening antimonopoles of the A1 configurations are located on two horizontal circles. 

Hence the first member of the A1 solution when $m=1\frac{1}{2}$ has a multimonopole of charge $+2\frac{1}{2}$ at $r=0$ and partially screened by two antimonopoles on the plane through the positive x-axis and the z-axis. The next member has $M_0=4$ and $M_{A}=-4$, and hence the net magnetic charge is zero. As $m$ increases in steps of one-half, $M_0$ increases by $1\frac{1}{2}$ monopole charge and the surrounding antimonopoles increases by two.

The subsequence 1s series has multimonopole charge of $M_0=(s+2)m-(s+1)s$, net magnetic charge of $M_\infty = s(s+1-m)$, and net antimonopole charge of $M_{A}=-2(s+1)(m-s)$, Table \ref{tab:1}. The A1s screening series possesses half-integer mutimonopole charge when the parameter $m$ is a half-odd-integer, whereas the B1s screening series possesses only integer values of multimonopole charge. The horizontal layers of screening antimonopoles increases as $s+1$. A more detail discussion of the higher 1s screening solutions is given in a separate work \cite{kn:13}.

\begin{table}[tbh]
\caption{The 1s Series of Screening Solutions. Here $b=m-s$ where $s\in\{0, 1, 2, 3, \dots\}$. Each series starts with $m=s+\frac{1}{2}$ and $m$ increases in steps of one-half. When $s$ is odd, we have the A1s series and when $s$ is zero or even we have the B1s series.}
\bigskip
\label{tab:1}
\begin{tabular}{|c|c|c|c|c|} \hline
\bf{1s Series}  & $s$                   & $M_0$                 & $M_\infty$     &   $M_{A}$\\ \hline
B$1$            &  ~~~0                 & $2m$                  & $0$            &   $-2m$   \\ \hline
A$1$            &  ~~~1                 & $3m-(2\times 1)$      & $1(2-m)$       &   $-4(m-1)$\\ \hline
B$1_1$          &  ~~~2                 & $4m-(3\times 2)$      & $2(3-m)$       &   $-6(m-2)$\\ \hline
A$1_s$          &  $s$ odd integer      & $(s+2)m-(s+1)s$       & $s(s+1-m)$     &   $-2(s+1)(m-s)$\\ \hline
B$1_s$          &  $s$ even integer     & $(s+2)m-(s+1)s$       & $s(s+1-m)$     &  $-2(s+1)(m-s)$\\ \hline
\end{tabular}
\end{table}

\section{The Multimonopole and the Finitely Separated 1-Monopoles}
\subsection{The 2s Series Multimonopole}
In the 2s series of solutions there are totally no antimonopoles in the configurations. All the monopoles and multimonopole possess only positive topological magnetic charge. Similar to the 1s series, we farther subdivide this series into the A2s solutions when $s$ is even, and the B2s solutions when $s$ is odd. The boundary conditions of the 2s series of solutions are
\begin{eqnarray}
\psi(r)|_{r\rightarrow \infty} &\rightarrow& -m,~~\psi(r)|_{r\rightarrow 0}\rightarrow (m+1);~~~
G(\theta, 0) = G(\theta, 2\pi)=0; \nonumber\\
R(\theta)\sin\theta|_{\theta\rightarrow 0} &\rightarrow& (m+s),~~R(\theta)\sin\theta|_{\theta\rightarrow \pi}\rightarrow -(m+s). 
\label{eq.28}
\end{eqnarray}
Similar to the 1s series, the difference between the A2s and B2s solutions lies in the profile function, $R(\theta)$. For the A2s solutions, $R(\theta)\cos\theta|_{\theta=\pi/2} = 1$, whereas for the B2s solutions, $R(\frac{\pi}{2}) = 0$.

The A2s series of solutions are configurations with a single multimonopole at the origin. The A2 solutions \cite{kn:9} is the first member of the A2s series and when $m=-\frac{1}{2}$, the multimonopole charge is $2\frac{1}{2}$. In fact, the A1 and the A2 solutions actually converge into a single multimonopole solution when $m=-\frac{1}{2}$ in both solutions. Hence half-integer topological charge multimonopoles do exit in the SU(2) YMH theory.

The B2 solutions \cite{kn:9} are the finitely separated 1-monopoles solutions. This series starts with $m=\frac{1}{2}$, where the solution is a three 1-monopoles configuration. As the parameter $m$ increases in steps of one-half, the number of monopoles in the configuration increases by one. Hence in the B2 solutions, we can have odd as well as even numbers of finitely separated 1-monopoles. In Ref.\cite{kn:9}, we reported on configurations with only even numbers of finitely separated monopoles in the B2 solutions.

The higher series of A2s and B2s solutions when $s=3, 4, 5, \dots$, start with higher multimonopole charge. Some properties of the 2s series of solutions are tabulated in table \ref{tab:2}.

\begin{table}[tbh]
\caption{The 2s Series of Multimonopole Solutions. Here $b=m+s$, where $s\in\{1, 2, 3, \dots\}$, and $M_A=0$. For even $s$, we have the A2s series where $m=-\frac{1}{2}, 0, \frac{1}{2}, 1, \frac{3}{2}, 2, \frac{5}{2}, \dots$ and for the odd $s$, we have the B2s series where $m=\frac{1}{2}, 1, \frac{3}{2}, 2, \frac{5}{2}, 3, \dots$.}
\bigskip
\label{tab:2}
\begin{tabular}{|c|c|c|c|c|} \hline
\bf{2s Series}  & $s$                   & $M_0$       & $M_\infty$   &  \bf{Configuration}\\ \hline
B$2$            &  ~~~1                 & $0$         & $2(m+1)$     &   $2(m+1)$ 1-monopoles\\ \hline
A$2$            &  ~~~2                 & $m+3$       & $m+3$        &   $(m+3)$-monopole\\ \hline
B$2_s$          &  $s$ odd integer      & $0$         & $2(m+s)$     &   $2(m+s)$ 1-monopoles\\ \hline
A$2_s$          &  $s$ even integer     & $m+s+1$     & $m+s+1$      &  $(m+s+1)$-monopole\\ \hline
\end{tabular}
\end{table}

\subsection{The C Series Multimonopole}
The C solution is a series of multimonopole solutions with half-integer topological magnetic charge \cite{kn:12}. The multimonopole is located at the origin, $r=0$, and has positive topological magnetic charge, $M = m \in \{\frac{1}{2}, 1, \frac{3}{2}, 2, \dots\}$. This series of solutions is solved by writing, $p=0$ and $b=m$ in Eq.(\ref{eq.13}) to (\ref{eq.15}). The solutions obtained are
\begin{eqnarray}
\psi(r)=\frac{1}{1+r},~~~R(\theta)= m\csc\theta,~~~G(\theta,\phi)=m\csc\theta \tan(m\phi).
\label{eq.29}
\end{eqnarray}
The boundary conditions are $\psi(r)|_{r\rightarrow 0}=1, ~\psi(r)|_{r\rightarrow \infty}=0$; $R(\theta)|_{\theta\rightarrow 0,\pi}\rightarrow +\infty$; and $G(\theta,0)=G(\theta,2\pi)=0$. 

The magnetic charge of the monopole at $r=0$ is calculated to be one-half of the normal t'Hooft-Polyakov monopole charge when $m=\frac{1}{2}$. This half-monopole solution of the C series possesses only mirror symmetry at the vertical plane through the x-z axes and a Dirac-like string singularity along the negative z-axis. 

When $m=0$ and 1, the C configurations possess topological magnetic charge one. The 1-monopole when, $m=0$, is just the radially symmetric Wu-Yang type 1-monopole \cite{kn:9} whereas the 1-monopole when, $m=1$,  possesses only mirror symmetry at the vertical x-z plane. A 3-D surface plot of the Abelian magnetic energy density, $B_iB_i$, at small $r$  reveals that this particular 1-monopole is actually made up of two half-monopoles. More detail discussions on half-monopoles can be found in Ref.\cite{kn:12} and \cite{kn:14}.

The C series of solutions continue to support higher topological magnetic charge multimonople as $m$ increases in steps of one-half. Hence when $m=0, \frac{1}{2}, 1, \frac{3}{2}, 2, \frac{5}{2}, \dots $, the topological charge of the multimonopole is $M= 1, \frac{1}{2}, 1, \frac{3}{2}, 2, \frac{5}{2}, \dots $ respectively. All these C multimonopoles except for the case when, $m=0$, which is radially symmetrical, seem to be made up of half-monopoles \cite{kn:12}.

\section{The Anti-Configurations}
In this section, we wish to show that for each possible monopole solution of the magnetic ansatz (\ref{eq.10}), there exist a monopole configuration where the directions of the magnetic field, $B_i$, of the original configuration are reversed. The reversing of the directions of the magnetic field implies that monopoles become antimonopoles and vice versa. This configuration where monopoles and antimonopoles change signs, we called the anti-configuration.
The anti-configurations can be constructed with some modifications of the magnetic ansatz (\ref{eq.10}). The static gauge field potentials and the Higgs field which will lead to these new anti-solutions are given respectively by 
\begin{eqnarray}
A_{\mu}^a &=& -\frac{1}{r}\psi(r)\left(\hat{u}^{a}_{\theta}\hat{\phi}_\mu + \hat{u}^{a}_{\phi}\hat{\theta}_{\mu}\right)  
+ \frac{1}{r}R(\theta)\left(\hat{u}^{a}_{\phi}\hat{r}_{\mu} + \hat{u}^{a}_{r}\hat{\phi}_{\mu}\right)\nonumber\\
&+& \frac{1}{r}G(\theta,\phi)\left(\hat{u}^{a}_{r}\hat{\theta}_\mu - \hat{u}^{a}_{\theta}\hat{r}_{\mu}\right),\nonumber\\
\Phi^a &=& \Phi_{1}~\hat{u}^{a}_{r} + \Phi_{2}~\hat{u}^{a}_{\theta} + \Phi_3~\hat{u}^{a}_{\phi}.
\label{eq.30}
\end{eqnarray}

\noindent where $\Phi_1 = \frac{1}{r}\psi(r), ~\Phi_2 = \frac{1}{r}R(\theta), ~\Phi_3 = \frac{1}{r}G(\theta, \phi)$. The spherical coordinate orthonormal unit vectors, $\hat{u}^{a}_{r}$, $\hat{u}^{a}_{\theta}$, and $\hat{u}^{a}_{\phi}$ with $\phi$ winding number $n$ are defined by 
\begin{eqnarray}
\hat{u}^{a}_{r} &=& \sin\theta ~\cos n\phi ~\delta^a_1 + \sin\theta ~\sin n\phi ~\delta^a_2 + \cos\theta \delta^a_3,\nonumber\\
\hat{u}^{a}_{\theta} &=& \cos\theta ~\cos n\phi ~\delta^a_1 + \cos\theta ~\sin n\phi ~\delta^a_2 - \sin\theta ~\delta^a_3,\nonumber\\
\hat{u}^{a}_{\phi} &=& -\sin n\phi ~\delta^a_1 + \cos n\phi ~\delta^a_2,
\label{eq.31}
\end{eqnarray}

\noindent where $r=\sqrt{x^i x_i}$, ~$\theta=\cos^{-1}(x_3/r)$, and~$\phi=\tan^{-1}(x_2/x_1)$. The unit vectors, $\hat{u}^{a}_{r}$, $\hat{u}^{a}_{\theta}$, and $\hat{u}^{a}_{\phi}$ will reduced to the normal spherical coordinate unit vectors $\hat{r}^a$, $\hat{\theta}^a$, and $\hat{\phi}^a$ when $n=1$. The gauge field strength tensor and the covariant derivative of the Higgs field are respectively
\begin{eqnarray}
F^a_{\mu\nu} &=& \frac{1}{r^2}\hat{u}^{a}_{r}\left\{n(\dot{R} + R \cot\theta + 2\psi - \psi^2) + G^{\phi} \csc\theta\right\}(\hat{\phi}_{\mu}\hat{\theta}_{\nu} - \hat{\phi}_{\nu}\hat{\theta}_{\mu})\nonumber\\
&+& \frac{1}{r^2}\left\{\hat{u}^{a}_{\theta} ~nR(1-\psi) + \hat{u}^{a}_{\phi} ~nG(1-\psi)\right\}(\hat{\phi}_{\mu}\hat{\theta}_{\nu} - \hat{\phi}_{\nu}\hat{\theta}_{\mu})\nonumber\\ 
&+& \frac{1}{r^2}\left\{\hat{u}^{a}_{r}~nR(1-\psi) + \hat{u}^{a}_{\phi} ~nG(\cot\theta-R)\right\}(\hat{r}_{\mu}\hat{\phi}_{\nu} - \hat{r}_{\nu}\hat{\phi}_{\mu})\nonumber\\
&+& \frac{1}{r^2}\hat{u}^{a}_{\theta}\left\{n(r\psi^{\prime} + R \cot\theta - R^{2}) + G^\phi \csc\theta\right\}(\hat{r}_{\mu}\hat{\phi}_{\nu} - \hat{r}_{\nu}\hat{\phi}_{\mu})\nonumber\\ 
&+& \frac{1}{r^2}\left\{-\hat{u}^{a}_{r}G(1-\psi) + \hat{u}^{a}_{\theta}(\dot{G} + RG)\right\}(\hat{r}_{\mu}\hat{\theta}_{\nu} - \hat{r}_{\nu}\hat{\theta}_{\mu})\nonumber\\
&+& \frac{1}{r^2}\left\{- \hat{u}^{a}_{\phi}(r\psi^{\prime} + \dot{R} - G^2)\right\}(\hat{r}_{\mu}\hat{\theta}_{\nu} - \hat{r}_{\nu}\hat{\theta}_{\mu}),\nonumber\\
D_{\mu}\Phi^{a} &=& \frac{1}{r^2}\left\{\hat{u}^{a}_{r}~(r\psi^{\prime}-\psi-R^2-G^2) - \hat{u}^{a}_{\theta}~R(1-\psi) - \hat{u}^{a}_{\phi}~G(1-\psi)\right\}\hat{r}_{\mu}\nonumber\\
&+& \frac{1}{r^2}\left\{-\hat{u}^{a}_{r}~R(1-\psi) + \hat{u}^{a}_{\theta}~(\dot{R}+\psi-\psi^2-G^2) + \hat{u}^{a}_{\phi}~(\dot{G}+RG)\right\}\hat{\theta}_{\mu}\nonumber\\
&+& \frac{1}{r^2}\left\{-\hat{u}^{a}_{r}~nG(1-\psi) - \hat{u}^{a}_{\theta}~nG(\cot\theta-R)\right\}\hat{\phi}_{\mu}\nonumber\\
&+& \frac{1}{r^2}\left\{\hat{u}^{a}_{\phi}\left(~n(\psi-\psi^2+R\cot\theta-R^2)+G^\phi\csc\theta\right)\right\}\hat{\phi}_{\mu}.
\label{eq.32}
\end{eqnarray}

\noindent The gauge used is the radiation or Coulomb gauge, $\partial^i A^a_i = 0$, $A^a_0 = 0$. 
The ansatz (\ref{eq.30}) is substituted into the equations of motion (\ref{eq.3}) as well as the Bogomol'nyi equations (\ref{eq.9}) with the negative sign. Upon choosing the $\phi$ winding number $n=-1$ these equations can be simplified to just four first order differential equations,
\begin{equation}
r\psi^{\prime} + \psi - \psi^2 = -p,
\label{eq.33}
\end{equation}
\begin{equation}
\dot{R} + R\cot\theta - R^2 = p - b^2\csc^2\theta,
\label{eq.34}
\end{equation}
\begin{equation}
\dot{G} + G\cot\theta = 0,~~~G^\phi\csc\theta + G^2 = b^2\csc^2\theta.
\label{eq.35}
\end{equation}
Hence any original monopole solutions, $\psi(r)$, $R(\theta)$, and $G(\theta,\phi)$ of the ansatz (\ref{eq.10}) are also solutions of the anti-configuration ansatz (\ref{eq.30}) with the profile function $G(\theta,\phi)=-b\csc\theta\tan(b\phi)$.

Similar to the ansatz (\ref{eq.10}), the ansatz (\ref{eq.30})  also has $A_{\mu} = \hat{\Phi}^{a}A^{a}_{\mu} = 0$. Hence the Abelian electric field is zero and the Abelian magnetic field is independent of the gauge fields $A^a_\mu$. The Higgs field of Eq.(\ref{eq.30}) can be written from the spherical coordinate system to the Cartesian coordinate system, 
\begin{eqnarray}
\Phi^a &=& \Phi_{1}~\hat{u}^{a}_{r} + \Phi_{2}~\hat{u}^{a}_{\theta} + \Phi_3~\hat{u}^{a}_{\phi}\nonumber\\
&=& \bar{\Phi}_1 ~\delta^{a1} + \bar{\Phi}_2 ~\delta^{a2} + \bar{\Phi}_3 ~\delta^{a3},
\label{eq.36}
\end{eqnarray}
\begin{eqnarray}
\mbox{where}~~~\bar{\Phi}_1 &=& \sin\theta \cos n\phi ~\Phi_1 + \cos\theta \cos n\phi ~\Phi_2 - \sin n\phi ~\Phi_3
= |\Phi|\cos\alpha \sin\beta\nonumber\\
\bar{\Phi}_2 &=& \sin\theta \sin n\phi ~\Phi_1 + \cos\theta \sin n\phi ~\Phi_2 + \cos n\phi ~\Phi_3
= |\Phi|\cos\alpha \cos\beta\nonumber\\
\bar{\Phi}_3 &=& \cos\theta ~\Phi_1 - \sin\theta ~\Phi_2 = |\Phi|\sin\alpha,
\label{eq.37}
\end{eqnarray}
and the Higgs unit vector is then simplified to 
\begin{eqnarray}
\hat{\Phi}^a = \cos\alpha \sin\beta ~\delta^{a1} + \cos\alpha \cos\beta ~\delta^{a2} + \sin\alpha ~\delta^{a3}.
\label{eq.38}
\end{eqnarray}
Hence the Abelian magnetic field, $B_i$, is similar in expression to Eq.(\ref{eq.24}) with $\beta = \gamma - n\phi = \gamma+\phi$ and $\sin\alpha$ and $\gamma$ given by Eq.(\ref{eq.23}). For a particular solution, $\psi$, $R$, and $G$ of ansatz (\ref{eq.10}), the solution of ansatz (\ref{eq.30}) is $\psi$, $R$, and $-G$. Therefore the directions of the Abelian magnetic field of the anti-configuration are reversed.
The Abelian field magnetic flux and the topological magnetic charge $M$ are as given in Eq.(\ref{eq.25}) and Eq.(\ref{eq.26}) respectively. Hence the magnetic flux and the topological magnetic charge are also reversed in sign.

\section{Comments}
In Ref.\cite{kn:9}, we have reported on configurations of even number of finitely separated monopoles in the B2 solutions. By noticing that $m$ can take half-integer values with the function $G$ still a single value function, the B2 solutions here can contain both odd and even numbers of finitely separated monopoles, starting with three 1-monopoles configuration when $m=\frac{1}{2}$. Its anti-configuration is shown in Fig.(\ref{fig.2}). Similarly, the number of B1 solutions is doubled when $m$ takes half-integer values and the B1 configuration when $m=\frac{1}{2}$ is just a pair of monopole and antimonopole. When $m=\frac{3}{2}$, the anti-configuration has three monopoles surrounding the 3-antimonopole, Fig.(\ref{fig.1}). No half-integer multimonopole is found in the B series of solutions.

Half-integer topological charge multimonopoles are recorded in the A and C series of monopoles solutions. 
We have also notice the screening monopoles/antimonopole are always of unit topological charge. Detail studies of the axially symmetric monopole solutions \cite{kn:11}, half-monopole solutions \cite{kn:12}, and higher series 1s solutions  \cite{kn:13}, are reported in separate works.

We have found that the SU(2) YMH theory does support monopole of one-half topological charge \cite{kn:12}, \cite{kn:14}, as well as multimonopole of half-integer topological charge. Also a 1-monopole need not be radially symmetrical as the 1-monopole of the C solution when $m=1$, possesses only mirror symmetry and is actually made up of two one-half monopoles.

The profile functions, $\psi(r)$, and $G(\theta,\phi)$ of Eq.(\ref{eq.16}) are standard solutions for all the monopoles solutions of the ansatz (\ref{eq.10}). It is the profile function, $R(\theta)$, that determines the different types of monopoles configurations of the gauge field potentials (\ref{eq.10}). Table 3 summarizes the properties of the different series of monopoles configurations of the ansatz (\ref{eq.10}). These series depend on the behavior of $R(\theta)$ at $\theta=0$, $\frac{\pi}{2}$, and $\pi$ rad.

\begin{table}[tbh]
\caption{A Table of Summary for the Different Monopoles Series of Solutions. The parameter $p$ is zero for the C solutions  and $p=m(m+1)$ for the other series of solutions. $m$ increases in steps of one-half up each series and $s=0, 1, 2, 3, \dots$.} 
\bigskip
\label{tab:3}
\begin{tabular}{|c|c|c|c|c|c|c|} \hline
\bf{Series}& $s$     & m                 & b & $R(\theta)\sin\theta|_{\theta\rightarrow 0}$ & $R(\theta)\sin\theta|_{\theta\rightarrow \pi}$ & $R(\frac{\pi}{2})$ \\ \hline
A1s        & odd     & $\geq s+\frac{1}{2}$ & $m-s$ & $-(m-s)$ & $m-s$ & $R(\theta)\cos\theta|_{\theta\rightarrow \frac{\pi}{2}}$=1 \\ \hline
B1s        & 0, even & $\geq s+\frac{1}{2}$ & $m-s$ & $-(m-s)$ & $m-s$ & $0$    \\ \hline
A2s        & even    & $\geq -\frac{1}{2}$  & $m+s$ & $m+s$ & $-(m+s)$ & $R(\theta)\cos\theta|_{\theta\rightarrow \frac{\pi}{2}}=1$  \\ \hline
B2s        & odd     & $\geq\frac{1}{2}$     & $m+s$ & $m+s$ & $-(m+s)$ & $0$    \\ \hline
C          & 0       & $\geq\frac{1}{2}$     & $m$   & $m$   & $m$        & $m$  \\ \hline
\end{tabular}
\end{table}

The positions of the antimonopoles of the 1s screening solutions and the monopoles of the B2s series about the origin are determined by the zeros of $\psi(r)$, $R(\theta)$, and $G(\theta,\phi)$.

In section 6, we have proved that for every monopoles solutions of ansatz (\ref{eq.10}), there always exist an anti-configuration of this solution where the directions of its Abelian magnetic field and hence its topological magnetic charge sign are reversed. 

The Abelian magnetic field of the anti-B1 solution when $m=\frac{3}{2}$, anti-B2 solution when $m=\frac{1}{2}$, and anti-A2 solution when $m=-\frac{1}{2}$, are shown in Fig.(\ref{fig.1}), Fig.(\ref{fig.2}), and Fig.(\ref{fig.3}) respectively. A point plot of the topological magnetic charges $M_\infty$ versus $M_0$ is shown in Fig.(\ref{fig.4}) for the A1, A2, B1, B2, C, AS series of solutions and their anti-configurations.

\section{ACKNOWLEDGEMENTS}
The author, Rosy Teh, would like to thank Universiti Sains Malaysia for the short term research grant (Account No: 304/PFIZIK/634039).

\newpage

\section*{FIGURE CAPTIONS}
\begin{figure}[tbh]
\vspace{5in}
\hskip0.5in\special{bmp:anti-B1_m=1.5.jpg x=5in y=5in}
\caption{The magnetic field of the anti-B1 solution when $m=1\frac{1}{2}$. The 3-antimonopole at the origin of the coordinate axes is surrounded by three monopoles lying on the x-y plane.}
\label{fig.1}
\end{figure}

\begin{figure}[tbh]
\vspace{5in}
\hskip0.5in\special{bmp:anti-B2_m=0.5.jpg x=5in y=5in}
\caption{The magnetic field of the anti-B2 solution when $m=\frac{1}{2}$ has three finitely separated antimonopoles located on the x-y plane.}
\label{fig.2}
\end{figure}

\begin{figure}[tbh]
\vspace{5in}
\hskip0.5in\special{bmp:anti-A2_m=-0.5.jpg x=5in y=5in}
\caption{The magnetic field of the anti-A2 solution when $m=-\frac{1}{2}$, showing the $2\frac{1}{2}$-antimonopole at the origin of the coordinate axes.}
\label{fig.3}
\end{figure}

\begin{figure}[tbh]
\vspace{5.5in}
\hskip0.5in\special{bmp:pointplotABC.jpg x=5.5in y=5.5in}
\caption{A point plot of the net magnetic charges $M$ at large $r$ versus the magnetic charges $M_0$ at $r=0$ for the series of configurations and anti-configurations of A1, A2, B1, B2, C and AS.}
\label{fig.4}
\end{figure}
\end{document}